\documentclass{elsarticle}
\usepackage{epsfig}

\begin{document}

\title{Flavor Asymmetry of the Nucleon Sea and $W$-Boson Production}

\author[uiuc]{Ruizhe Yang}
\ead{yangrz@npl.uiuc.edu}

\author[uiuc]{Jen-Chieh Peng}
\ead{jcpeng@uiuc.edu}

\author[uiuc]{Matthias Gro\ss{}e-Perdekamp}
\ead{mgp@uiuc.edu}

\address[uiuc]{University of Illinois at Urbana-Champaign, Urbana, IL
  61801, United States}

\date{\today}

\begin{abstract}
  The advantage and feasibility of using $W$-boson production to
  extract unique information on the flavor asymmetry of the $\bar u$
  and $\bar d$ sea-quark distributions in the proton are examined. The
  $W^+$ and $W^-$ production cross section ratios in $p+p$ collisions
  are shown to be sensitive to the $\bar d/ \bar u$ ratios, and they
  are free from charge-symmetry-breaking and nuclear-binding
  effects. The feasibility for measuring these ratios at the RHIC and
  LHC proton-proton colliders, as well as the expected sensitivity to
  the $\bar d/ \bar u$ ratios, are also presented.
\end{abstract} 


\maketitle

The earliest parton models assumed that the proton sea was flavor
symmetric, even though proton's valence quark distributions were known
to be flavor asymmetric.  Inherent in this assumption is that the
content of the sea is independent of the valence quark's composition.
The assumption of sea-quark flavor symmetry was not based on any known
physics, and it remained to be tested by experiments.
Neutrino-induced charm production experiments~\cite{conrad98} provided
clear evidences that the strange-quark content of the nucleon is only
about half of the up or down sea quarks. This flavor asymmetry is
attributed to the much heavier mass for strange quark compared to the
up and down quarks.  The similarity between the masses of up and down
quarks suggests that the nucleon sea should be nearly up-down
symmetric.  However, it was pointed out that the existence of a pion
cloud in the proton could lead to an asymmetric up-down
sea~\cite{thomas83}.

A measurement of the Gottfried integral in deep-inelastic scattering
(DIS) provides a direct check of the $\bar d/ \bar u$ flavor-symmetry
assumption. The Gottfried integral~\cite{gott67} is defined as
\begin{equation}
I_G = \int_0^1 \left[F^p_2 (x) - F^n_2 (x)\right]/x~ dx =
{1\over 3}+{2\over 3}\int_0^1 \left[\bar u_p(x)-\bar d_p(x)\right]dx,
\label{eq1}
\end{equation}
where $F^p_2$ and $F^n_2$ are the proton and neutron structure
functions measured in DIS experiments and $x$ is the fraction of the
nucleon's momentum carried by the quark. The second step in
Eq.~\ref{eq1} follows from the assumption of charge symmetry (CS) at
the partonic level, namely, $u_p(x)=d_n(x),~d_p(x)= u_n(x),~ \bar
u_p(x) = \bar d_n(x),$ and $ \bar u_p(x) = \bar d_n(x)$.  Under the
assumption of a symmetric sea, $\bar u_p$ = $\bar d_p$, the Gottfried
Sum Rule (GSR), $I_G = 1/3$, is obtained.  The most accurate test of
the GSR was reported by the New Muon Collaboration (NMC)~\cite{nmc91},
which measured $F^p_2$ and $F^n_2$ over the region $0.004 \le x \le
0.8$. They determined the Gottfried integral to be $ 0.235\pm 0.026$,
significantly below 1/3. This surprising result has generated much
interest. Although the violation of the GSR can be explained by
assuming unusual behavior of the parton distributions at very small
$x$, a more natural explanation is that the assumption $\bar u = \bar
d$ is invalid.

The proton-induced Drell-Yan (DY) process provides an independent
means to probe the flavor asymmetry of the nucleon sea~\cite{es}.  An
important advantage of the DY process is that the $x$ dependence of
$\bar d / \bar u$ asymmetry can be determined.  Using a 450 GeV proton
beam, the NA51 Collaboration~\cite{na51} at CERN measured dimuons
produced in $p+p$ and $p+d$ reaction and obtained $\bar u/\bar d =
0.51 \pm 0.04 (stat) \pm 0.05 (syst)$ at $x = 0.18$ and $\langle
M_{\mu\mu}\rangle = 5.22$ GeV.  At Fermilab, a DY experiment
(E866/NuSea) covering a broad kinematic range with high statistics has
been carried out~\cite{e866,peng,towell}. The E866 Collaboration
measured the DY cross section ratios for $p + d$ to that of $p + p$ at
the forward-rapidity region using intense 800 GeV proton beams. At
forward rapidity region and assuming the validity of charge symmetry,
one obtains

\begin{equation}
\sigma_{DY}(p+d)/2\sigma_{DY}(p+p) \simeq
(1+\bar d(x)/\bar u(x))/2.
\label{eq:2}
\end{equation}

\noindent This ratio was found to be significantly different from
unity for $0.015 <x< 0.35$, indicating an excess of $\bar d$ with
respect to $\bar u$ over an appreciable range in $x$.

The HERMES Collaboration has also reported a semi-inclusive DIS
measurement of charged pions from hydrogen and deuterium
targets~\cite{hermes}.  Based on the differences between charged-pion
yields from the two targets, $\bar d - \bar u$ is determined in the
kinematic range $0.02 < x < 0.3$ and 1 GeV$^2$/c$^2 < Q^2 <$ 10
GeV$^2$/c$^2$. The HERMES results are consistent with the E866 results
obtained at significantly higher $Q^2$.

Many theoretical models, including meson-cloud model, chiral-quark
model, Pauli-blocking model, instanton model, chiral-quark soliton
model, and statistical model, have been proposed to explain the $\bar
d/ \bar u$ asymmetry, as reviewed in~\cite{kumano98,garvey02}.  While
these models can describe the general trend of the $\bar d / \bar u$
asymmetry, they all have difficulties explaining the $\bar d / \bar u$
data at large $x$ ($x>0.2$)~\cite{melnitchouk}.  Since the
perturbative process gives a symmetric $\bar d/ \bar u$ while a
non-perturbative process is needed to generate an asymmetric $\bar d/
\bar u$ sea, the relative importance of these two components is
directly reflected in the $\bar d/ \bar u$ ratios. Thus, it would be
very important to have new measurements sensitive to the $\bar d /
\bar u$ ratios at $x>0.2$.  The upcoming Fermilab E906 Drell-Yan
experiment~\cite{e906} plans to extend the measurement to larger $x$
region.

With the advent of $p+p$ colliders at RHIC and LHC, an independent
technique to study the $\bar d/ \bar u$ asymmetry now becomes
available.  By measuring the ratio of $W^+$ versus $W^-$ production in
unpolarized $p+p$ collision, the $\bar d/ \bar u$ asymmetry can be
determined~\cite{doncheski94,peng95,bourrely94} with some distinct
advantages over the existing methods. First, this method does not
require the assumption of the validity of charge symmetry.  All
existing experimental evidences for $\bar d/ \bar u$ asymmetry depend
on the comparison between DIS or DY scattering cross sections off
hydrogen versus deuterium targets. The possibility that charge
symmetry could be violated at the parton level has been discussed by
several authors~\cite{ma1,ma2,sather,rodionov,benesh1,londergan2}.  Ma
and collaborators~\cite{ma1,ma2} pointed out that the violation of the
Gottfried Sum Rule can be caused by charge symmetry violation as well
as by flavor asymmetry of the nucleon sea. They also showed that DY
experiments, such as NA51 and E866, are subject to both flavor
asymmetry and charge symmetry violation effects. In fact, an even
larger amount of flavor asymmetry is required to compensate for the
possible charge symmetry violation effect~\cite{steffens96}. A
comparison between $W$ production in $p+p$ collision with the NA51 and
E866 Drell-Yan experiments would disentangle the flavor asymmetry from
the charge symmetry violation effects.

Another advantage of $W$ production in $p+p$ collision is that it is
free from any nuclear effects. As pointed out by several
authors~\cite{wally93,braun94,sawicki93,schmidt01}, the nuclear
modification of parton distributions should be taken into account for
DIS and DY process involving deuterium targets. The nuclear shadowing
effect for deuteron at small $x$ could lead to a 4\% to 10\% decrease
in the evaluation of the Gottfried integral by the
NMC~\cite{wally93,schmidt01}. Moreover, the nucleon Fermi motion at
large $x$ also affects the extraction of neutron structure function
and would cause additional uncertainty in the evaluation of the
Gottfried integral~\cite{braun94}. The nuclear effects and the
associated uncertainty are absent in $W$ production in $p+p$
production.

Finally, the $W$ production is sensitive to $\bar d/ \bar u$ flavor
asymmetry at a $Q^2$ scale of $\sim$ 6500 GeV$^2/$c$^2$, significantly
larger than all existing measurements. This offers the opportunity to
examine the QCD evolution of the sea-quark flavor asymmetry. The large
mass of $W$ also implied that the RHIC data are sensitive to the
sea-quark flavor asymmetry at the large $x$ region, which remains
poorly known both experimentally and theoretically as discussed
earlier.

The differential cross section for $W^+$ production in hadron-hadron
collision can be written as~\cite{barger87}

\begin{eqnarray}
{d \sigma \over dx_F} (W^+) & = & K {\sqrt 2 \pi \over 3} G_F
\left({x_1 x_2 \over {x_1 + x_2}}\right)
\left\{ \cos^2 \theta_c \,
[u(x_1) \bar d(x_2) + \bar d(x_1) u(x_2)] + \right. \nonumber \\
&  & \left. \sin^2 \theta_c \,
[u(x_1) \bar s(x_2) + \bar s(x_1) u(x_2)] \right\},
\end{eqnarray}

\noindent where $u(x), d(x),$ and $s(x)$ signify the up, down, and
strange quark distribution functions in the hadrons.  $x_1, x_2$ are
the fractional momenta carried by the partons in the colliding hadron
pair and $x_F = x_1 - x_2$.  $G_F$ is Fermi coupling constant and
$\theta_c$ is the Cabbibo angle.  The factor $K$ takes into account
the contributions from first-order QCD corrections~\cite{barger87}

\begin{eqnarray}
K \simeq 1 + {8\pi \over 9} \alpha_s(Q^2).
\end{eqnarray}

\noindent At the $W$ mass scale, $\alpha_s \simeq 0.1158$ and $K
\simeq 1.323$.  This indicates that higher-order QCD processes are
relatively unimportant for $W$ production.  An analogous expression
for $W^-$ production cross section is given as

\begin{eqnarray}
{d \sigma \over dx_F} (W^-) & = & K {\sqrt 2 \pi \over 3} G_F
\left({x_1 x_2 \over {x_1 + x_2}}\right)
\left\{ \cos^2 \theta_c \,
[\bar u(x_1) d(x_2) +  d(x_1) \bar u(x_2)] + \right. \nonumber \\
&  & \left. \sin^2 \theta_c \,
[\bar u(x_1) s(x_2) + s(x_1) \bar u(x_2)] \right\},
\end{eqnarray}

An interesting quantity to be considered is the ratio of the
differential cross sections for $W^+$ and $W^-$ production.  If one
ignores the much smaller contribution from the strange quarks, this
ratio can be written as

\begin{eqnarray}
R(x_F) \equiv {{{d \sigma \over dx_F} (W^+)} \over
{{d \sigma \over dx_F} (W^-)}} =
{{u(x_1) \bar d(x_2) +  \bar d(x_1) u(x_2)} \over
{\bar u(x_1) d(x_2) +  d(x_1) \bar u(x_2)}}.
\end{eqnarray}

\noindent For $p + p$ collision, it is evident that $R(x_F)$ is
symmetric with respect to $x_F = 0$, namely, $R(x_F) = R(-x_F)$.  It
is clear that $R(x_F)$ in $p+p$ collision is sensitive to the
sea-quark distributions in the proton.  At large $x_F$, we have

\begin{eqnarray}
R(x_F \gg 0) =
{{u(x_1) \bar d(x_2) +  \bar d(x_1) u(x_2)} \over
{\bar u(x_1) d(x_2) +  d(x_1) \bar u(x_2)}} \approx
{u(x_1) \over d(x_1)} {\bar d(x_2) \over \bar u(x_2)}.
\end{eqnarray}

\noindent At $x_F = 0$, where $x_1 = x_2 = x$, one obtains

\begin{eqnarray}
R(x_F = 0) =
{{u(x) \bar d(x) +  \bar d(x) u(x)} \over
{\bar u(x) d(x) +  d(x) \bar u(x)}} =
{u(x) \over d(x)} {\bar d(x) \over \bar u(x)}.
\end{eqnarray}

\noindent As the $u(x)/d(x)$ ratios are already well known, a
measurement of $R(x_F)$ in $p + p$ collision gives an accurate
determination of the ratio $\bar d(x)/\bar u(x)$.

Figure 1 shows the predictions of $R(x_F)$ for $p + p$ collision at
$\sqrt s=500$~GeV.  Four different structure function sets together
with the full expressions for $W^+,W^-$ production cross sections
given by Eqs. (3) and (5) have been used in the calculations.  The
first PDF used here is MRS S0'~\cite{martin93}.  It assumes symmetric
$\bar u$ and $\bar d$ distributions, therefore, according to Eq. (8),
$R(x_F) \simeq 2$ at $x_F = 0$ as shown in Fig.~1.  The other three
PDFs used here allowed certain flavor asymmetry in nucleon sea.  New
experimental data from Drell-Yan measurement by E866 Collaboration is
included in the global fit performed by CTEQ6~\cite{pumplin02},
GJR08~\cite{gluck08} and MSTW2008~\cite{martin09} to determine
$x$-dependence of $\bar u$, $\bar d$ asymmetry in the nucleon sea.
Thus $R(x_F)$ for those three PDF are similar at $x_F = 0$ and are
significantly higher than 2 obtained in the MRS S0' case.
Table~\ref{t:rhicx} shows the $x_1$ and $x_2$ values for $W$
production at RHIC with center of mass energy 500 GeV.

\begin{table}[htbp]
\begin{center}
\begin{tabular}{c|ccccccccc}
\hline
$x_F$ &     0.0 &     0.1 &     0.2 &     0.3 &     0.4 &     0.5 &     0.6 &     0.7 &     0.8 \\ 
\hline 
$x_1$ &    0.16 &    0.22 &    0.29 &    0.37 &    0.46 &    0.55 &    0.64 &    0.73 &    0.83 \\ 
\hline 
$x_2$ &    0.16 &    0.12 &    0.09 &    0.07 &    0.06 &    0.05 &    0.04 &    0.03 &    0.03 \\ 
\hline
\end{tabular}
\caption{values for $x_1$ and $x_2$ at different $x_F$ for $W$ production
at $\sqrt{s} =$ 500 GeV.}
\label{t:rhicx}
\end{center}
\end{table}

Although Fig.~1 shows that the differences between the predictions of
$R(x_F)$ for various PDFs are quite conspicuous, in practice it is not
the $x_F$ distributions of the $W$ which are measured but rather the
charged leptons from the decay of the $W$-bosons.  The measured lepton
ratio is defined as:
\begin{eqnarray}
R(y_l)
&=& { d\sigma/dy_l (W^+ \to l^+)
\over d\sigma/dy_l (W^- \to l^-)},
\end{eqnarray}
where the lepton rapidity $y_l = 1/2 \ln \left[(E_l + p_l)/(E_l -
  p_l)\right]$ is defined in terms of the decay lepton's energy $E_l$
and longitudinal momentum $p_l$ in the laboratory frame.  The
differential cross section $d\sigma/dy_l$ is obtained by convoluting
the $q\overline q \rightarrow W$ cross section for each $x_F$ with the
relevant $W \rightarrow l\ \nu$ decay distribution,
$d\sigma/d\cos\theta \propto (1 \pm \cos\theta)^2$, where $\theta$ is
the angle between the lepton $l^{\pm}$ direction and the $W^{\pm}$
polarization in the $W$ rest frame.

In Fig.~2 we show the predicted lepton ratios $R(y_l)$ calculated for
various PDFs. The statistical uncertainties for the lepton ratios are
estimated for recorded luminosity of 300 pb$^{-1}$ at
RHIC~\cite{rhic}.  The acceptance is for PHENIX
experiment~\cite{phenix}, which covers $|y| < 0.35$ in central
rapidities and $-2.2 < y < -1.1$, $1.1 < y < 2.4$ in forward
rapidities.  Fig.~2 has clearly demonstrated that a measurement of
$R(y_l)$ at RHIC is able to distinguish flavor symmetric and flavor
asymmetric nucleon sea.

The calculation for $R(x_F)$ and $R(y_l)$ has also been carried out
for CMS experiment at LHC~\cite{cms}.  Fig.~3 shows results for
$R(x_F)$ at LHC.  At $x_F = 0$, all PDFs used here obtain similar
results for $R(x_F)$.  This is due to the fact that at much higher
c.m.s. energy, this measurement probes sea quark flavor asymmetry at
even lower $x$ compared to previous measurements from Drell-Yan
process and semi-inclusive DIS, and all four PDFs used here have
predicted that flavor asymmetry will diminish as $x \to 0$.
Table~\ref{t:lhcx} shows the $x_1$ and $x_2$ values for $W$ production
at LHC with center of mass energy 14 TeV.

\begin{table}[htbp]
\begin{center}
\begin{tabular}{c|ccccccc}
\hline
$x_F$ &     0.0 &     0.1 &     0.2 &     0.3 &     0.4 &     0.5 &     0.6 \\ 
\hline 
$x_1$ &    0.01 &    0.10 &    0.20 &    0.30 &    0.40 &    0.50 &    0.60 \\ 
\hline 
$x_2$ &  0.0114 &  0.0013 &  0.0007 &  0.0004 &  0.0003 &  0.0003 &  0.0002 \\
\hline
\end{tabular}
\caption{values for $x_1$ and $x_2$ at different $x_F$ for $W$ production
at $\sqrt{s} =$ 7 TeV.}
\label{t:lhcx}
\end{center}
\end{table}

Fig.~4 shows results of the lepton ratio $R(y_l)$ where integrated
luminosity is assumed to be 10 fb $^{-1}$ corresponding to one year
low luminosity running of $p + p$ collisions at $\sqrt s = 14$ TeV and
the pseudorapidity coverage is taken as $|\eta| < 5$~\cite{cms}.  The
sensitivity of $R(y_l)$ in Fig.~4 is more than sufficient to
differentiate flavor symmetry and asymmetry used in different
parameterizations.

In conclusion, $W$ production at RHIC and LHC would offer an
independent means to examine the $\bar d/\bar u$ flavor asymmetry in
the proton. Measurements of the cross section ratios of $W^+ \to l^+$
and $W^- \to l^-$ production in $p + p$ collisions would provide a
sensitive test of current PDFs. The $W$ production experiments at RHIC
and LHC will offer the unique opportunity of extracting the $\bar d/
\bar u$ flavor asymmetry at large $x$ and very high $Q^2$ without the
complications associated with the charge symmetry breaking effect and
nuclear binding effect. The proposed measurements are within the
capabilities of the existing detectors at RHIC and LHC and can be
carried out in the near future.

\newpage

\begin{figure}
\epsfig{figure=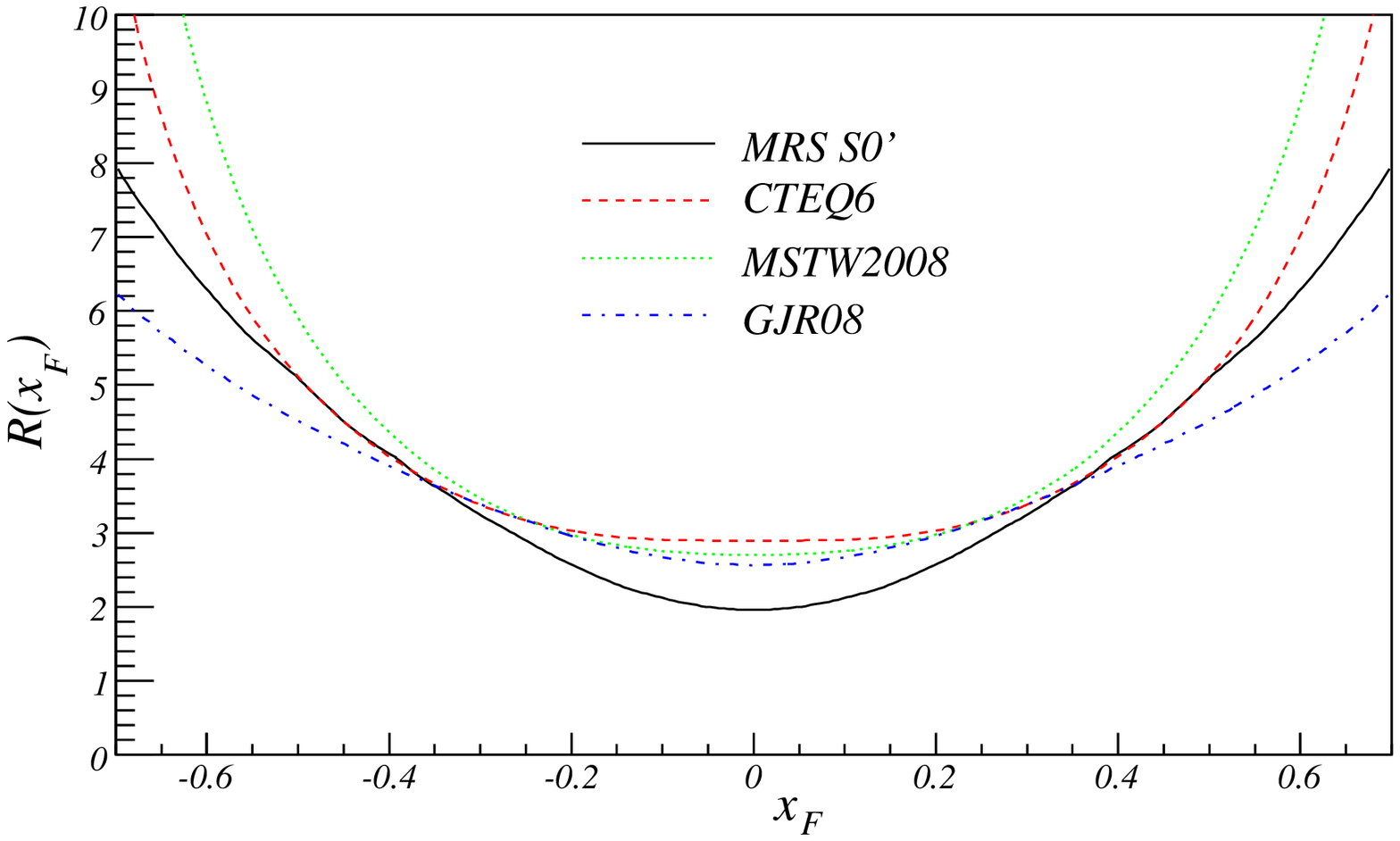,height=7.5cm}
\caption{Prediction of the ratio $R(x_F)$ as a function of $x_F$ for
p+p collision at $\sqrt s$ of 500 GeV using
various parton distribution functions.}
\end{figure}

\begin{figure}
\epsfig{figure=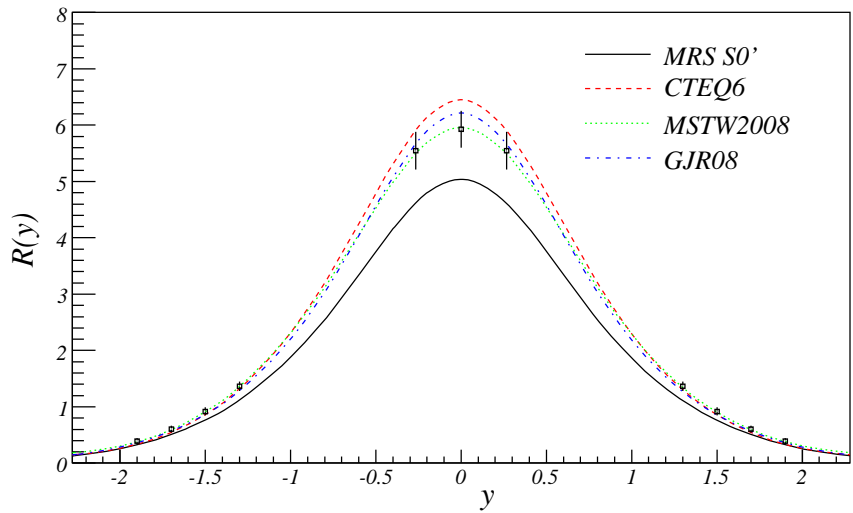,height=7.5cm}
\caption{Prediction of the ratio $R(y_l)$ as a function of $y$ for
p+p collision at $\sqrt s$ of 500 GeV using
various parton distribution functions. The projected sensitivities
for a run with recorded luminosity of 300 pb$^{-1}$ for the PHENIX detector
are also shown.}
\end{figure}

\begin{figure}
\epsfig{figure=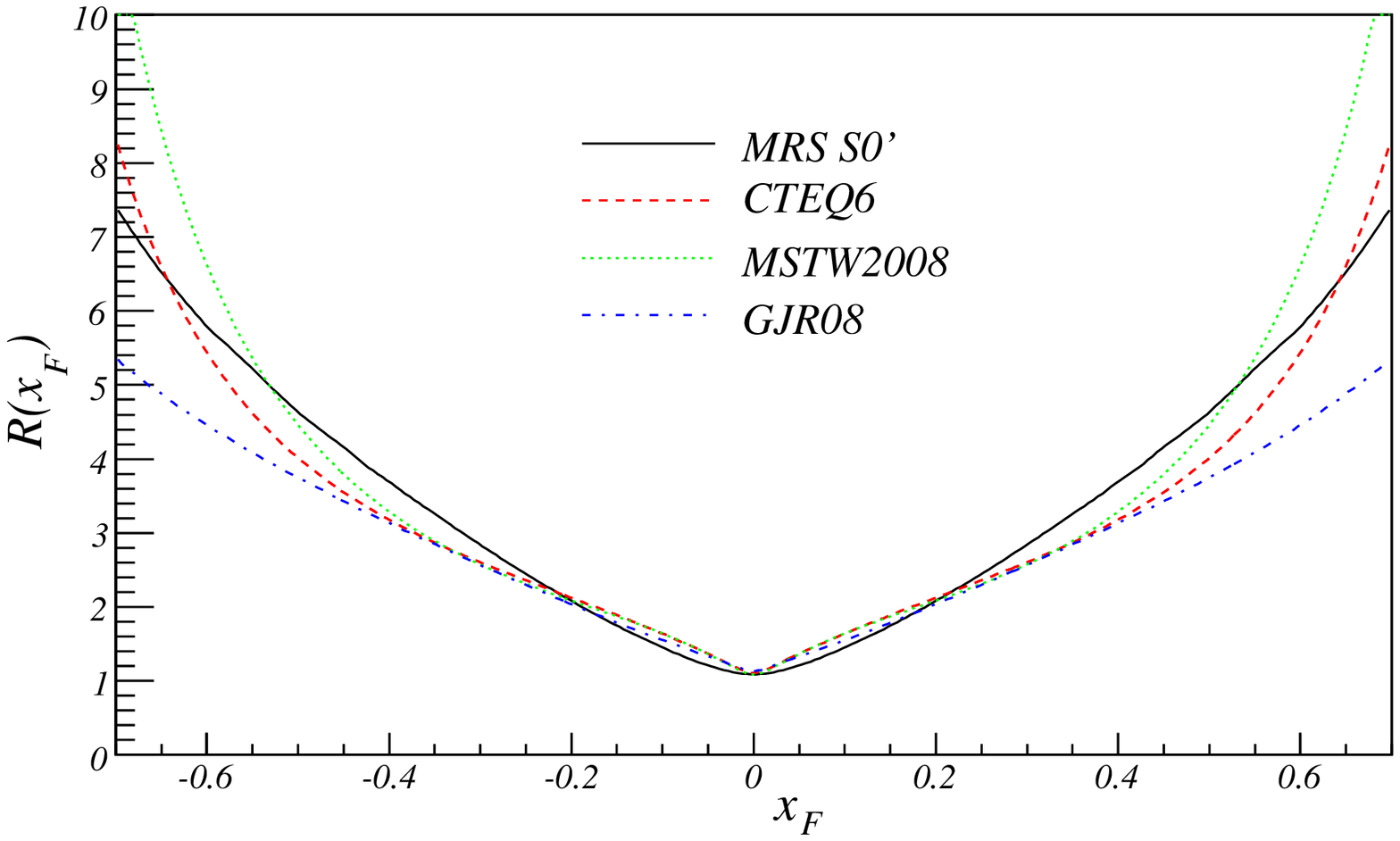,height=7.5cm}
\caption{Prediction of the ratio $R(x_F)$ as a function of $x_F$ for
p+p collision at $\sqrt s$ of 14 TeV using
various parton distribution functions.}
\end{figure}

\begin{figure}
\epsfig{figure=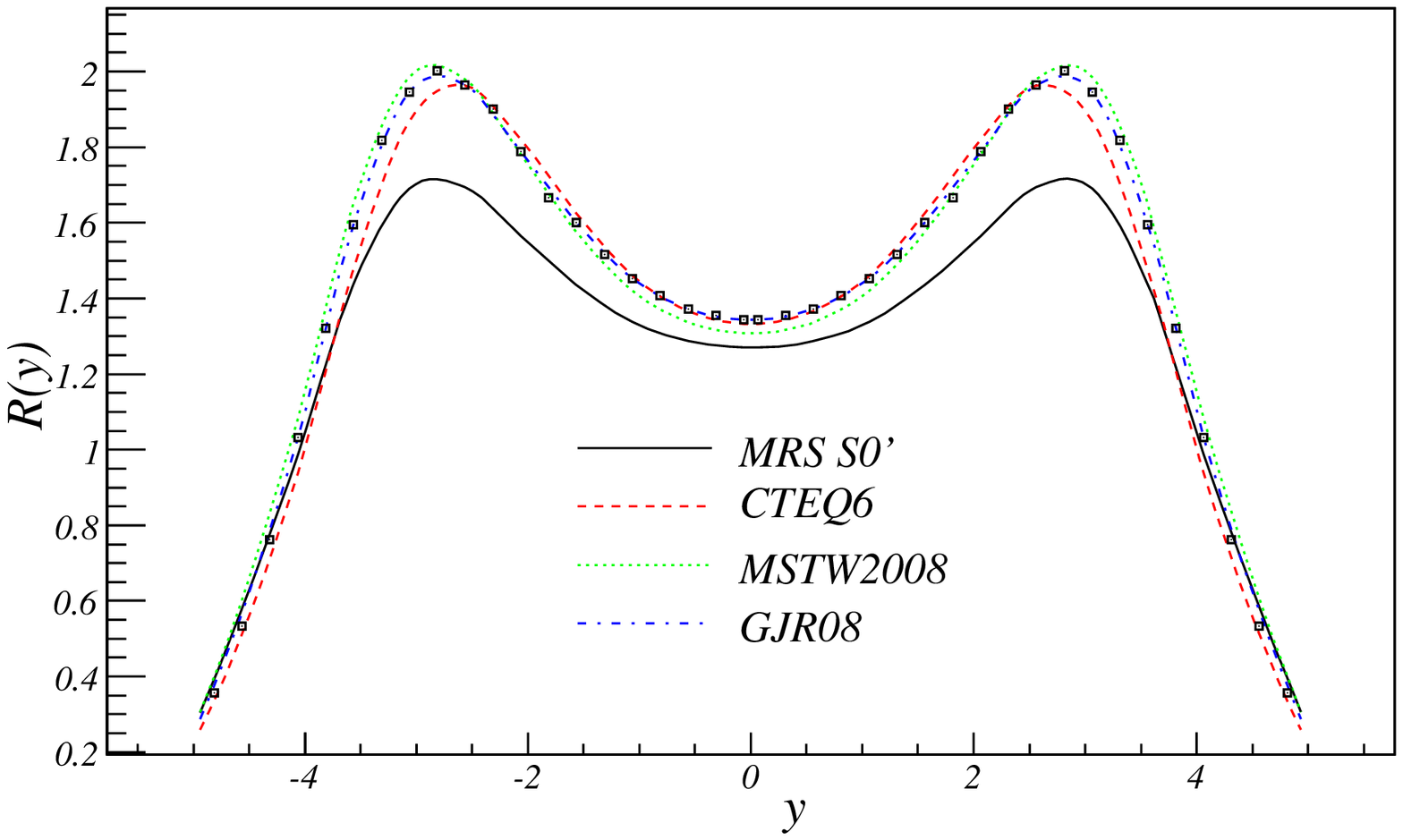,height=7.5cm}
\caption{Prediction of the ratio $R(y_l)$ as a function of $y$ for
p+p collision at $\sqrt s$ of 14 TeV using
various parton distribution functions. The projected sensitivities
for a run with integrated luminosity of 10 fb$^{-1}$ for the CMS detector
are also shown.}
\end{figure}


\begin{thebibliography}{100}
\bibitem{conrad98} J.M. Conrad, M.H. Shaevitz and T. Bolton, Rev. Mod. Phys.
{\bf 70} (1998) 1341. 
\bibitem{thomas83} A.W. Thomas, Phys. Lett. {\bf B126} (1983) 97.
\bibitem{gott67} K. Gottfried, Phys. Rev. Lett. {\bf 18} (1967) 1174.
\bibitem {nmc91} P. Amaudruz {\it et al.}, Phys. Rev. Lett. {\bf 66} (1991) 
2712; M. Arneodo {\it et al.}, Phys. Rev. {\bf D55} (1994) R1.
\bibitem{es} S.D. Ellis and W.J. Stirling, Phys. Lett. {\bf B256} (1991) 258.
\bibitem{na51} A. Baldit {\it et al.}, Phys. Lett. {\bf B332} (1994) 244.
\bibitem{e866} E.H. Hawker {\it et al.},  Phys. Rev.  Lett. {\bf 80} 
(1998) 3715.
\bibitem{peng} J.C. Peng {\it et al.}, Phys. Rev. {\bf D58} (1998) 092004.
\bibitem{towell} R.S. Towell {\it et al.}, Phys. Rev. {\bf D64} (2001) 052002.
\bibitem{hermes} K. Ackerstaff  {\it et al.}, Phys. Rev. Lett. {\bf 81} 
(1998) 5519.
\bibitem{kumano98} S. Kumano, Phys. Rep. {\bf 303} (1998) 183.
\bibitem{garvey02} G.T. Garvey and J.C. Peng, Prog. Part. Nucl. Phys. {\bf 47} (
2001) 203.
\bibitem{e906} D.~Geesaman, P.~Reimer {\em et al.}, Fermilab Proposal P906, 1999, http://www.phy.anl.gov/mep/drell-yan/ .
\bibitem{melnitchouk} W. Melnitchouk, J. Speth and A.W. Thomas, Phys. Rev.
{\bf D59} (1999) 014033.
\bibitem{doncheski94} M.A. Doncheski {\it et al.}, Phys. Rev. 
{\bf D49} (1994) 3261.
\bibitem{peng95} J.C. Peng and D.M. Jansen, Phys. Lett. {\bf B354} (1995) 460.
\bibitem{bourrely94} C.~Bourrely and J.~Soffer, Nucl. Phys. {\bf B423} (1994) 329.
\bibitem{ma1} B.-Q. Ma, Phys. Lett. {\bf B274} (1992) 111.
\bibitem{ma2} B.-Q. Ma, A. Sch\"{a}fer and  W. Greiner, Phys. Rev. 
{\bf D47} (1993) 51.
\bibitem{sather} E. Sather, Phys. Lett. {\bf B274} (1992) 433.
\bibitem{rodionov} E.N. Rodionov, A.W. Thomas and J.T. Londergan, 
Int. J. Mod. Phys. Lett. {\bf A9} (1994) 1799.
\bibitem{benesh1} C.J. Benesh and T. Goldman, Phys. Rev. {\bf C55} (1997) 441.
\bibitem{londergan2} J.T. Londergan and A.W. Thomas,
Progress in Particle and Nuclear Physics, {\bf 41} (1998) 49.
\bibitem{steffens96} F.M. Steffens and A.W. Thomas, Phys. Lett. 
{\bf B389} (1996) 217.
\bibitem{wally93} W. Melnitchouk and A.W. Thomas, Phys. Rev.
{\bf D47} (1993) 3783.
\bibitem{braun94} M.A. Braun and M.V. Tokarev, Phys. Lett. {\bf B320} 
(1994) 381.
\bibitem{sawicki93} M. Sawicki and J.P. Vary, Phys. Rev. Lett. {\bf 71}
(1993) 1320.
\bibitem{schmidt01} I. Schmidt and J.-J. Yang, Eur. Phys. J. {\bf C20}
(2001) 63.
\bibitem{barger87} V.D. Barger and R.J.N. Phillips, Collider
Physics (Addison - Wesley Publishing Company, 1987).

\bibitem{martin93} A.D.~Martin, W.J.~Stirling, R.G.~Roberts, Phys. Lett. {\bf B306} (1993) 145.
\bibitem{pumplin02} J.~Pumplin, D.R.~Stump, J.~Huston, H.L.~Lai, P.M.~Nadolsky and W.K.~Tung, J. High Energy Phys. {\bf 0207} (2002) 012.
\bibitem{gluck08} M.~Gluck, P.~Jimenez-Delgado and E.~Reya, Eur.\ Phys.\ J.\  C {\bf 53} (2008) 355.
\bibitem{martin09} A.D.~Martin, W.J.~Stirling, R.S.~Thorne and G.~Watt, arXiv:0901.0002 [hep-ph].
\bibitem{rhic} G.~Bunce {\it et al.}, Plans for the RHIC Spin Physics Program, 2008, http://spin.riken.bnl.gov/rsc/report/spinplan\_2008/spinplan08.pdf .
\bibitem{phenix} K.~Adcox {\it et al.} (PHENIX Collaboration), Nucl. Inst. Meth. {\bf A499} (2003) 469.
\bibitem{cms} CMS Collaboration, Physics Technical Design Report Volume I: Detector Performance and Software, CERN/LHCC 2006-001, 2006.
\end{thebibliography}
\end{document}